%% file: iapichino-invisible.tex
\documentclass[
    final            
  ]
  {aipproc}

\usepackage{aas_macros} 
\layoutstyle{8x11single}
\usepackage{color}		
\usepackage{epsfig}

\begin{document}

\title{Turbulence modeling and the physics of the intra-cluster medium}


\author{L. Iapichino}{
  address={Zentrum f\"ur Astronomie der Universit\"at Heidelberg,
Institut f\"ur Theoretische Astrophysik, Albert-Ueberle-Str.~2, D-69120
Heidelberg, Germany}
}

\author{A. Maier}{
  address={Lehrstuhl f\"ur Astronomie, Universit\"at W\"urzburg,
Am Hubland, D-97074 W\"urzburg, Germany}
}

\author{W. Schmidt}{
  address={Institut f\"ur Astrophysik, Universit\"at
  G\"ottingen, Friedrich-Hund-Platz 1, D-37077 G\"ottingen, Germany}
}

\author{J. C. Niemeyer}{
  address={Institut f\"ur Astrophysik, Universit\"at
  G\"ottingen, Friedrich-Hund-Platz 1, D-37077 G\"ottingen, Germany}
}

\begin{abstract}
FEARLESS (Fluid mEchanics with Adaptively Refined Large Eddy SimulationS) is a new numerical scheme arising from the combined use of subgrid scale (SGS) model for turbulence at the unresolved length scales and adaptive mesh refinement (AMR) for resolving the large scales. This tool is
especially suitable for the study of turbulent flows in strongly clumped media. In this contribution, the main features of FEARLESS are briefly outlined. We then summarize the main results of FEARLESS cosmological simulations of galaxy cluster evolution. In clusters, the production of turbulence is
closely correlated with merger events; for minor mergers, we find that turbulent dissipation affects the cluster energy budget only locally. The level of entropy in the cluster core is enhanced in FEARLESS simulations, in accord with a better modeling of the unresolved flow, and with its feedback on the resolved mixing in the ICM. 
\end{abstract}

\maketitle

\section{Introduction}

The formation and evolution of the large scale structure, which proceeds through the process of hierarchical clustering, plays a key role for the conversion of potential gravitational energy into kinetic and internal energy in galaxy clusters and groups. In this process, cluster mergers induce bulk motions in the intra-cluster medium (henceforth ICM), 
with velocities of the order of $1000\ \mathrm{km\ s^{-1}}$. The shearing instabilities associated with the merger events and the ubiquitous shock waves in the ICM are thus expected to stir the gas and make the flow turbulent.

Cluster formation is a typical example of turbulence generation in a strongly clumped medium. As the Eulerian approach is generally considered to be superior to the Lagrangian one in modeling hydrodynamical instabilities \cite{ams07,tbm08,mmb09}), grid-based numerical codes are a suitable tool to address this problem, especially when the Adaptive Mesh Refinement (henceforth AMR, \cite{Berger1984,bc89,n05}) technique is used. 
On the other hand, as in most of the astrophysical fluids, the Reynolds number in the ICM is estimated to be $Re \gg 1$ (although with many issues on the value of the kinematic viscosity; \cite{nm01,snb03,fsc03,rmf05,j07}), implying a number of degrees of freedom of the dynamical system \cite{ll87,snh06} which is not manageable even in state-of-the-art numerical simulations\footnote{In case of cosmological simulations, the problem is even more severe because the evolution of galaxy clusters has to be followed in the cosmological context of their collapse, i.e. in simulated computational domains much larger than the cluster scale itself.}.
Therefore, even with AMR, it is usually not possible to resolve all the turbulent cascade, down to the dissipative length scale \cite{snh06}.

In many fields of computational fluid dynamics, the influence of unresolved turbulence on the resolved scales is modeled by means of heuristic subgrid-scale (SGS) models, coupled to the large scales of the system, for which the fluid equations are solved (Large Eddy Simulations, LES; \cite{Lesieur1996}). LES are customary, for example, in simulations of Type Ia supernovae \cite{nh95,rhn02,rh05,snhr06,rhs07}. Recently, \cite{sb08} developed a similar tool designed for studying Rayleigh-Taylor driven turbulence in the interaction between AGN outflows and the ICM \cite{bsh09,bs09}.

In this contribution, we describe a novel numerical tool that unites LES and AMR, called FEARLESS (Fluid mEchanics with Adaptively Refined Large Eddy SimulationS). FEARLESS combines the adaptive refinement of the regions where turbulent flows develop with a consistent modeling of the SGS turbulence, and is argued to be a suitable tool in simulations of turbulent clumped flows. The numerical formalism and its first application to the physics of galaxy clusters is presented in \cite{mis09}. The main features and results are collected here.

\section{Numerical tools}

FEARLESS has been implemented into the ENZO code\footnote{ENZO
 homepage: \texttt{http://lca.ucsd.edu/software/enzo/}}, v.~1.0 \cite{obb05}, an AMR, grid-based hybrid (hydrodynamics plus N-Body) code based on 
the PPM solver \cite{wc84} modified for the study of
cosmology \cite{bns95}, which provides the necessary infrastructure for performing numerical simulations of cosmological structure formation.

\subsection{Adaptive mesh refinement of turbulent flows}

As a first step in the development of FEARLESS, some work has been devoted to the development of refinement criteria which are best suited for refining turbulent flows. One such criterion was introduced in \cite{sfh09}, and is based on the regional variability of structural invariants of the flow, i.e.~variables related to the spatial derivatives of the flow velocity. An example is the 
modulus of the vorticity $\vec{\omega} = \nabla \times \vec{v}$ (the
curl of the velocity field), expected to become high in regions where the flow is turbulent.  The regional threshold for triggering the refinement is expressed in the comparison of the cell value of the variable $q(\vec{x},t)$ and the average and the standard deviation of $q$, calculated on a local grid patch:

\begin{equation}
\label{h0973-local}
q(\vec{x},t) \ge \langle q \rangle_i(t) + \alpha \lambda_i(t)
\end{equation}
where $\lambda_i$ is the maximum between the average $\langle q \rangle $ and the standard deviation of $q$ in the grid patch $i$, and $\alpha$ is a tunable parameter.

This technique has been profitably used in simulations of idealized minor mergers \cite{ias08} and in full cosmological simulations of galaxy clusters, focused on minor merger events \cite{in08}. In both cases, the new AMR criteria provide a better resolution of the turbulent flow, as demonstrated by larger values of the velocity dispersion on resolved scales and (for \cite{in08}) by an increased floor of entropy in the cluster core, due to an increased mixing. 

The new criteria have been used also in a recent study of major mergers \cite{pim09}, allowing to follow the evolution of merger shocks and the related post-shock region. The evolution of turbulence in that region and in the cluster core has been explored, and interesting morphological comparisons with the observed symmetrical radio relics in the merging cluster Abell 3376 \cite{bdn06} have been performed.

\subsection{Subgrid scale model}

Here, we give a general and qualitative account of the main features of the SGS model; we refer to \cite{snh06} and \cite{mis09} for a more detailed and rigorous description.


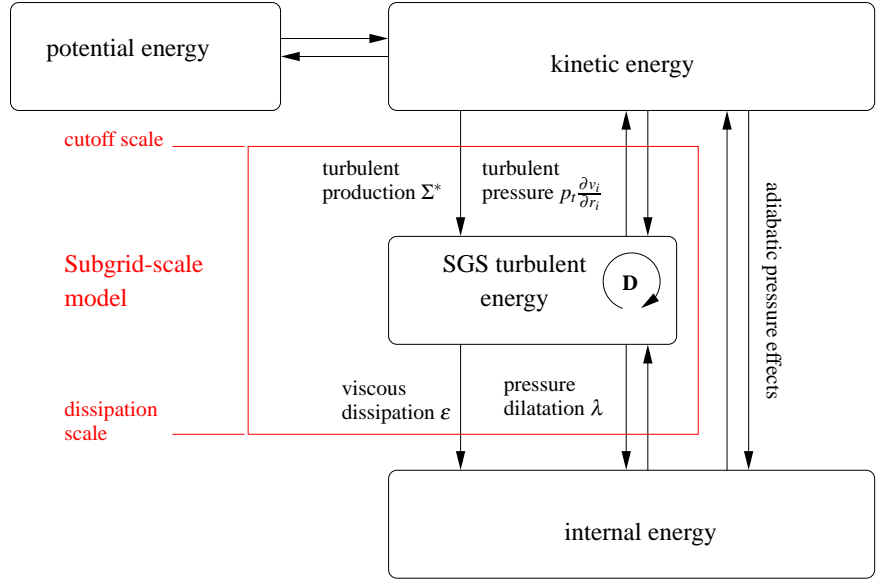
\begin{figure}
\centering
\resizebox{0.7\textwidth}{!}{
\input{energyscheme3b.pstex_t}}
\caption{Graphical description of the energy budget in FEARLESS, showing the  energy components and the exchange terms between them. From \cite{maier08}.}
\label{h0973-fig1}
\end{figure}

As shown by \cite{snh06}, the equations of motion for a compressible, viscous, self-gravitating fluid can be
decomposed into large-scale (resolved) and small-scale (unresolved)
parts, using the Germano filter formalism \cite{Germano1992}.
By means of filtering, any field quantity $a$ can be split into a smoothed part
$\langle a \rangle$ and a fluctuating part $a'$, where $\langle a \rangle$
varies only at scales greater than the prescribed filter length.

Following this procedure, one can define the SGS turbulent energy $e_{\mathrm{t}}$, which can be interpreted as a subgrid energy buffer between the resolved kinetic energy and the internal energy (graphically shown in Fig.~\ref{h0973-fig1}). The turbulent energy is governed by an equation of the following form:
\begin{equation}
\frac{\partial}{\partial t} \langle \rho \rangle e_{\mathrm{t}}+\frac{\partial}{\partial r_j}\hat{v}_j \langle \rho \rangle e_{\mathrm{t}}=\ D +\Sigma+ \Gamma - \langle \rho \rangle (\lambda+\epsilon)\,\,,
\label{h0973-etsum}
\end{equation}
where $\rho$ is the gas density and $\hat{v}_j$ is the density-weighted filtered velocity component, according to Favre \cite{Favre1969}. The quantities on the right-hand side of Eq.~\ref{h0973-etsum} determine the evolution of $e_{\mathrm{t}}$ and are the turbulent diffusion term $D$, the turbulent production term $\Sigma$, the pressure dilatation term $\lambda$ and the viscous dissipation term $\epsilon$. Their role in the energy budget is visually described in Fig.~\ref{h0973-fig1}, and the way they are modelled (i.e.~their so-called closures) represent the core of the SGS model. The filtering procedure couples the hydrodynamical equations to the unresolved scales by the inclusion of terms from Eq.~\ref{h0973-etsum}, as described in \cite{snh06,mis09}.

\subsection{Combining AMR and LES}

A severe limitation of the SGS model is the implicit assumption of the isotropy of the velocity fluctuations on subgrid scales. This limits the LES methodology to flows where all anisotropies stemming from large scale features, like boundary conditions or external forces, can be resolved. In FEARLESS, this issue is cured by the adaptive local adjustment of the grid resolution, in order to ensure that the anisotropic, energy-containing scales are resolved everywhere as much as possible. In this way, it is assumed that turbulence is asymptotically isotropic on length scales comparable to or less than the grid resolution. 

When a grid location is flagged for refinement in Enzo, a new finer grid is created, 
and the cell values on the finer grid are generated by interpolating them from the coarser grid 
using a conservative interpolation scheme. At each timestep of the coarse grid, the values from the fine grid are averaged and the values computed on the 
coarse grid (in the region where fine and coarse grid overlap) are replaced. An additional step of the refinement procedure in FEARLESS is a scale correction in the energy budget, resulting in a energy transfer from the SGS turbulent to the resolved kinetic energy (or the opposite for grid derefinement). The relation between the SGS turbulent energies at different levels of refinement $l_{1}$ and $l_{2}$ is given by the assumption of Kolmogorov scaling \cite{k41,f95}: 
\begin{equation}
\frac{e_{\mathrm{t},1}}{e_{\mathrm{t},2}}\sim
\left(\frac{l_{1}}{l_{2}}\right)^{2/3}\,\,.
\label{eq:qscale}
\end{equation}

The required correction steps for the resolved velocity components and for the turbulent velocity in the newly created fine grid (the primed quantities in the following equations) are: 
\begin{equation}
\hat{v}'_i =\hat{v}_i \sqrt{1+\frac{e_{\mathrm{t}}}{\hat{e}_{\mathrm{kin}}}\left(1-r_{\Delta}^{-2/3}\right)},
\end{equation}

\begin{equation}
e'_{\mathrm{t}} = e_{\mathrm{t}} r_{\Delta}^{-2/3}
\end{equation}
where $\hat{e}_{\mathrm{kin}}$ is the resolved kinetic energy and
$r_{\Delta}$ is the refinement factor of the mesh. The resolved 
energy is adjusted such that the sum of resolved energy and turbulent
energy remains conserved. The opposite procedure is applied at grid derefinement.

\section{Galaxy cluster simulations with FEARLESS}

The numerical technique presented in the last section has been applied to cosmological simulations of the formation and evolution of a galaxy cluster \cite{mis09}. Two cluster simulations, run with and without the use of the SGS model but otherwise with an identical numerical setup, are compared. 
 
A flat $\mathrm{\Lambda}$CDM background cosmology is assumed, with parameters
$\Omega_{\mathrm \Lambda} = 0.7$, $\Omega_{\mathrm m} = 0.3$,
$\Omega_{\mathrm b} = 0.04$, $h = 0.7$, $\sigma_8 = 0.9$, and $n =
1$. The simulations have been initialized at redshift $z_{\mathrm{in}}
= 60$ using the transfer function by \cite{eh99}, and then
evolved to $z = 0$. 

The computational box has a size of $128\ \mathrm{Mpc}\ h^{-1}$. The root grid (level $l=0$) has $128^3$ cells and $128^3$ N-body
particles. A static child grid ($l=1$) is nested inside the root grid with a
size of $64\ \mathrm{Mpc}\ h^{-1}$, $128^3$ cells and $128^3$ N-body particles. The mass of each
particle in this grid is $9 \times 10^9\, M_{\odot}\ h^{-1}$. Inside this grid,
AMR is allowed in a volume of $38.4\, \mathrm{Mpc}\ h^{-1}$, using the overdensity refinement criterion as
described in \cite{in08} with an overdensity factor $f=4.0$. The refinement
factor between two levels was set to 2, allowing for an effective resolution
of $7.8\, \mathrm{kpc}\ h^{-1}$. 

The static and dynamically refined grids were nested at the place of
formation of a galaxy cluster, previously identified in a low-resolution run using the HOP algorithm \cite{eh98}.  The cluster did not experience in its recent history ($z < 1$) any major merger, therefore it was considered an ideal case to study the role of minor mergers for injecting turbulent energy in the ICM.

\begin{figure}
  \resizebox{0.85\textwidth}{!}{\includegraphics{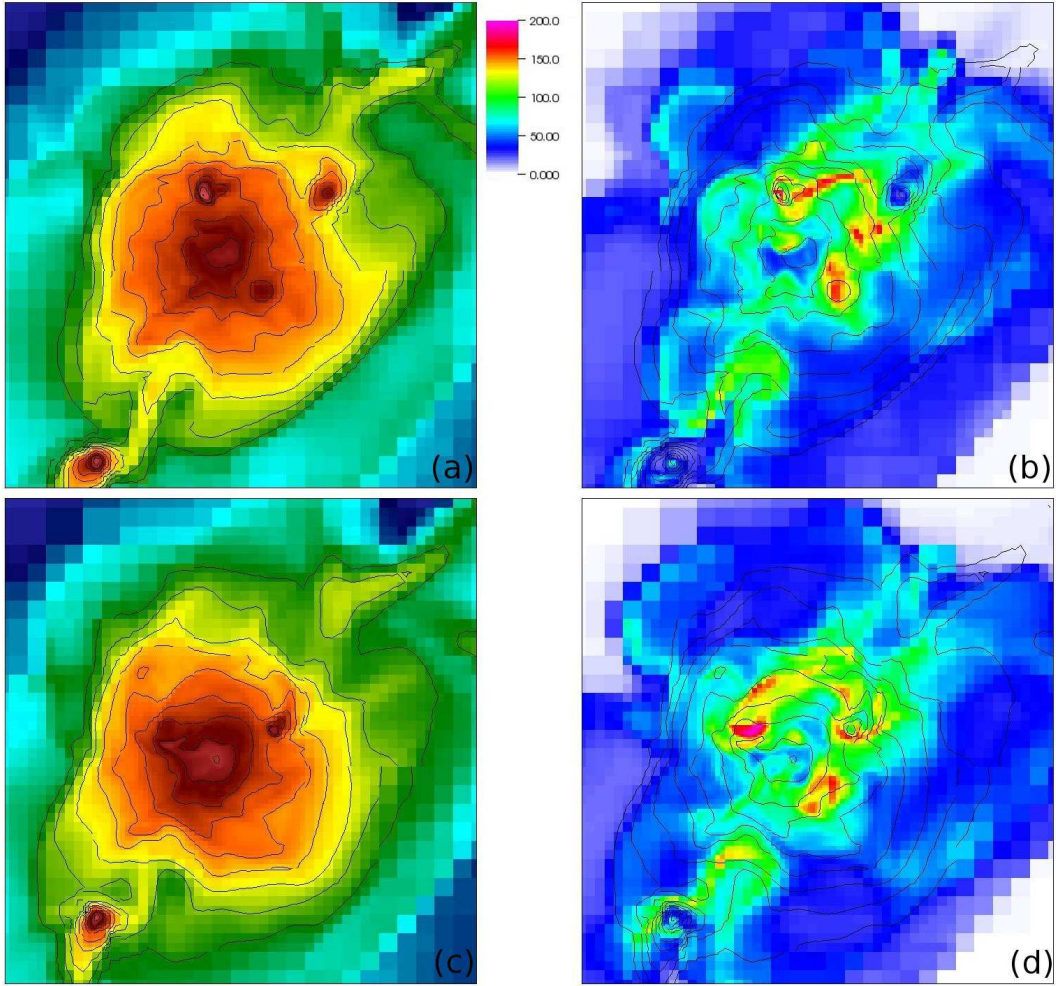}}
  \caption{Slices of baryon density (left-hand panels, $a$ and $c$) and
  turbulent velocity $q = \sqrt{2 e_{\mathrm{t}}}$ (right-hand panels, $b$ and $d$) at
different redshifts $z$, for the cosmological simulation run with FEARLESS. The density is
logarithmically color coded, whereas $q$ is linearly coded
in $\mathrm{km\ s^{-1}}$, according to the colorbar on the left of panel $b$. The overlayed contours show density. 
The slices show a region of $6.4 \times 6.4\, \mathrm{Mpc}\ h^{-1}$
around the center of the main cluster followed in the
simulation. Panels $a$ and $b$ refer to $z = 0.05$, panels $c$ and $d$
to $z = 0$.}
  \label{h0973-fig3}
\end{figure}

In Fig.~\ref{h0973-fig3}, a short time
series of density and SGS turbulent velocity slices is presented. Several minor mergers in the ICM can be identified, especially at the locations where the SGS turbulence is large. Indeed, a considerable amount of turbulent energy is localized in front and in the wake of the merging clumps
(panels \ref{h0973-fig3}$b$ and $d$).  The morphological evolution in Fig.~\ref{h0973-fig3} provides therefore a remarkable insight on the markedly local behavior of the production and dissipation of turbulence in clusters.

The energy content of the turbulent flow in the ICM is rather small, in agreement with the subsonic nature of the flow. The SGS turbulent energy is generally smaller than $1 \%$ of the internal energy, while the energy associated with the velocity dispersion at the length scale of $100\ \mathrm{kpc}\ h^{-1}$ is at the level of a few percent. On the other hand, the turbulent contribution to the energy budget is more relevant at the location and in the wake of the merging subclumps. According to \cite{mis09}, the value of the SGS turbulent energy in the vicinity of a subclump is one order of magnitude larger than the typical average value in the ICM. In addition to the numerical dissipation of the code, which converts  kinetic energy directly into internal energy, there is an additional dissipation channel due to the energy flux from resolved scales to the 
internal energy through the SGS energy buffer.

\begin{figure}
  \resizebox{\textwidth}{!}{\includegraphics{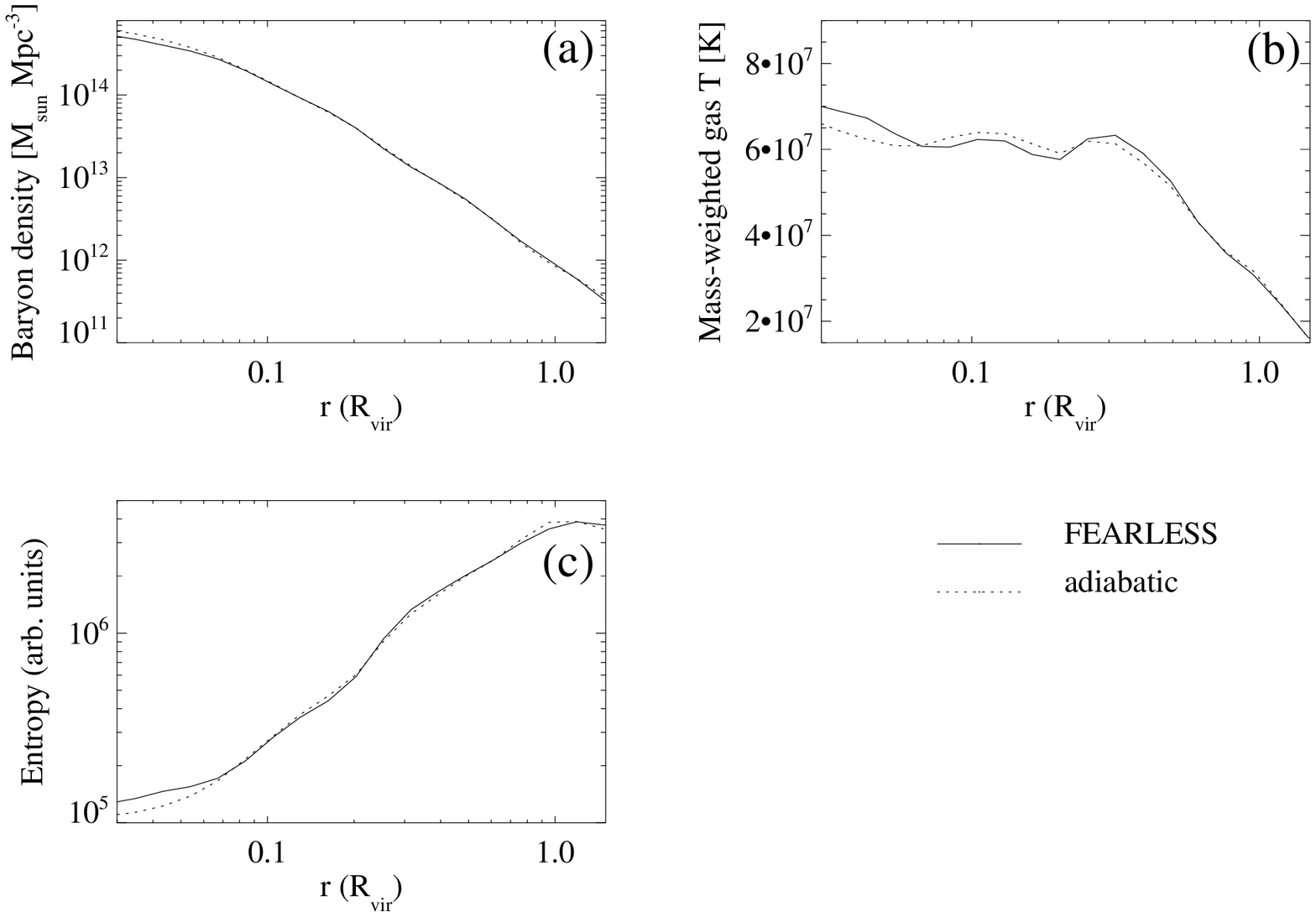}}
  \caption{Cluster radial profiles of selected quantities at $z = 0$. The
  dotted line refers to the simulation without SGS model, whereas the
  solid line is for the simulation using FEARLESS. The radii on the $x$-axis are expressed in fraction of the virial radius, $R_{\mathrm{vir}}=1.37\ \mathrm{Mpc}\ h^{-1}$ for the considered cluster. Panel $a$ reports the gas density, in $b$ is the mass-weighted temperature and in $c$ the gas entropy, defined in the text.}
  \label{radials}
\end{figure}

An interesting result, closely linked with the point just discussed above, comes from the comparison between radial
profiles in the FEARLESS simulation and in the standard
adiabatic run (Fig.~\ref{radials}). A change in the
temperature and density profiles at the cluster center is clearly visible. In particular, in the FEARLESS run $T$ is larger ($3\%$) for central distances $r < 0.07\
R_{\mathrm{vir}}$, with respect to the standard run (Fig.~\ref{radials}$b$). Consequently, the gas in the core is less
dense (Fig.~\ref{radials}$a$), so that the ICM locally remains in hydrostatic equilibrium. The local
energy budget in the cluster core is therefore modified by the SGS
model. 

The effects of the SGS model is reflected 
in the entropy which is defined, as customary in cluster physics, as 
\begin{equation}
K=\frac{T}{\rho^{\gamma-1}}
\end{equation}
where $\gamma = 5/3$ and $\rho$ is the gas density. The entropy in the cluster core is higher in the FEARLESS run as compared to the standard run
(Fig.~\ref{radials}$c$). This result is produced by the combination of two factors: the enhanced local dissipation of SGS turbulent to internal energy, provided
by the SGS model, and a moderate feedback of the SGS model on the resolved scales, increasing the magnitude of the resolved mixing (cf.~\cite{mis09} for more details).  

The enhanced entropy level in the core is in agreement with known results of the comparisons between SPH and grid codes on this issue 
 \cite{dvb05,wvc08,koc09}. Whenever the modeling of the turbulent flow in cluster simulations is improved, either by using grid codes rather than SPH, or by especially designed AMR criteria for refining turbulence \cite{in08}, or by using a SGS model \cite{mis09}, the level of core entropy is increased. Recently, \cite{s09} pointed out that the core temperature
and entropy in grid-based codes are affected by a spurious increase,
caused by the N-body noise in the gravitational force field. 
In our opinion, further investigations are certainly needed to disentangle physical from numerical effects on this issue.

\section{Summary and conclusions}

Large Eddy Simulations (LES) are based on the notion of 
filtering the fluid dynamic equations at a specific length scale. 
A separation between the resolved and the unresolved length scales of the
flow is thus performed. The latter are treated by means of a subgrid scale model, which is coupled to the hydrodynamical equations governing the
former. In principle, a single scale separation is not practical in simulations of clumped media, and moreover it is not immediately compatible with the concept of adaptive mesh refinement (AMR), often used to study astrophysical systems.
The solution proposed to this problem is the development of  a new numerical scheme exploiting AMR and LES in combination, called FEARLESS. 

In this contribution we briefly reviewed the basic ideas of the FEARLESS approach (AMR, SGS model and their coupling), and then the application of this technique to galaxy cluster simulations was discussed. The subsonic flow in the ICM are do not have a dominant role in the energy budget (as confirmed by previous analyses, e.g.~\cite{nb99,dvb05,cfv07,vbk09}), and therefore the impact of the SGS energy content is globally negligible. Nevertheless, at the location of the merging subclumps, the role of turbulent dissipation become sizeable. Such local effects are particularly relevant in the cluster core, where an enhanced entropy floor results in the FEARLESS simulation.

Future work in this area will regard cosmological simulations of major mergers. These events are particularly interesting because the turbulent energy injected by them has a significant role in the energy budget, with respect to minor mergers. As a first step, a study which only makes use of AMR, triggered by the regional variability of the compression rate of the flow, has been performed \cite{pim09}. Furthermore, it will be also important to explore the turbulent flow not only in the ICM, but also in the cluster outskirts and in the cosmic web. The knowledge of the turbulent state of this gas is potentially important for shaping the emission and absorption lines associated with the Warm-Hot Intergalactic Medium (see \cite{bsd08} for a review). From a computational viewpoint, numerical improvements in the FEARLESS scheme are ongoing \cite{sf09}, in order to better account for complex flows with large density gradients.

\begin{theacknowledgments}
The Enzo code is developed by the Laboratory for Computational Astrophysics at
the University of California in San Diego (\texttt{http://lca.ucsd.edu}). 
The numerical simulations were carried out on the SGI Altix 4700 \emph{HLRB II} of the Leibniz Computing Centre in Garching (Germany). 
\end{theacknowledgments}

\bibliographystyle{aipproc}  
\bibliography{cluster-index}

\end{document}

%% file: energyscheme3b.pstex_t
\begin{picture}(0,0)%
\includegraphics{energyscheme3b.pstex}%
\end{picture}%
\setlength{\unitlength}{4144sp}%
\begingroup\makeatletter\ifx\SetFigFont\undefined%
\gdef\SetFigFont#1#2#3#4#5{%
  \reset@font\fontsize{#1}{#2pt}%
  \fontfamily{#3}\fontseries{#4}\fontshape{#5}%
  \selectfont}%
\fi\endgroup%
\begin{picture}(10824,7224)(1114,-7273)
\put(1576,-736){\makebox(0,0)[lb]{\smash{{\SetFigFont{20}{24.0}{\familydefault}{\mddefault}{\updefault}{\color[rgb]{0,0,0}potential energy}%
}}}}
\put(8776,-931){\makebox(0,0)[b]{\smash{{\SetFigFont{20}{24.0}{\familydefault}{\mddefault}{\updefault}{\color[rgb]{0,0,0}kinetic energy}%
}}}}
\put(9001,-6781){\makebox(0,0)[b]{\smash{{\SetFigFont{20}{24.0}{\familydefault}{\mddefault}{\updefault}{\color[rgb]{0,0,0}internal energy}%
}}}}
\put(1801,-3856){\makebox(0,0)[lb]{\smash{{\SetFigFont{20}{24.0}{\familydefault}{\mddefault}{\updefault}{\color[rgb]{1,0,0}model}%
}}}}
\put(1801,-3436){\makebox(0,0)[lb]{\smash{{\SetFigFont{20}{24.0}{\familydefault}{\mddefault}{\updefault}{\color[rgb]{1,0,0}Subgrid-scale}%
}}}}
\put(1801,-5236){\makebox(0,0)[lb]{\smash{{\SetFigFont{17}{20.4}{\familydefault}{\mddefault}{\updefault}{\color[rgb]{1,0,0}dissipation}%
}}}}
\put(1801,-5536){\makebox(0,0)[lb]{\smash{{\SetFigFont{17}{20.4}{\familydefault}{\mddefault}{\updefault}{\color[rgb]{1,0,0}scale}%
}}}}
\put(7426,-3856){\makebox(0,0)[b]{\smash{{\SetFigFont{20}{24.0}{\familydefault}{\mddefault}{\updefault}{\color[rgb]{0,0,0}energy}%
}}}}
\put(7426,-3436){\makebox(0,0)[b]{\smash{{\SetFigFont{20}{24.0}{\familydefault}{\mddefault}{\updefault}{\color[rgb]{0,0,0}SGS turbulent}%
}}}}
\put(5026,-2221){\makebox(0,0)[lb]{\smash{{\SetFigFont{17}{20.4}{\familydefault}{\mddefault}{\updefault}{\color[rgb]{0,0,0}turbulent}%
}}}}
\put(5026,-2521){\makebox(0,0)[lb]{\smash{{\SetFigFont{17}{20.4}{\familydefault}{\mddefault}{\updefault}{\color[rgb]{0,0,0}production $\Sigma^*$}%
}}}}
\put(5266,-4966){\makebox(0,0)[lb]{\smash{{\SetFigFont{17}{20.4}{\familydefault}{\mddefault}{\updefault}{\color[rgb]{0,0,0}viscous}%
}}}}
\put(5266,-5266){\makebox(0,0)[lb]{\smash{{\SetFigFont{17}{20.4}{\familydefault}{\mddefault}{\updefault}{\color[rgb]{0,0,0}dissipation $\epsilon$}%
}}}}
\put(7291,-4921){\makebox(0,0)[lb]{\smash{{\SetFigFont{17}{20.4}{\familydefault}{\mddefault}{\updefault}{\color[rgb]{0,0,0}pressure}%
}}}}
\put(7291,-5221){\makebox(0,0)[lb]{\smash{{\SetFigFont{17}{20.4}{\familydefault}{\mddefault}{\updefault}{\color[rgb]{0,0,0}dilatation $\lambda$}%
}}}}
\put(7021,-2221){\makebox(0,0)[lb]{\smash{{\SetFigFont{17}{20.4}{\familydefault}{\mddefault}{\updefault}{\color[rgb]{0,0,0}turbulent}%
}}}}
\put(7021,-2521){\makebox(0,0)[lb]{\smash{{\SetFigFont{17}{20.4}{\familydefault}{\mddefault}{\updefault}{\color[rgb]{0,0,0}pressure $p_t\frac{\partial v_i}{\partial r_i}$}%
}}}}
\put(8776,-3661){\makebox(0,0)[lb]{\smash{{\SetFigFont{17}{20.4}{\familydefault}{\mddefault}{\updefault}{\color[rgb]{0,0,0}$\mathbf{D}$}%
}}}}
\put(1801,-1861){\makebox(0,0)[lb]{\smash{{\SetFigFont{17}{20.4}{\familydefault}{\mddefault}{\updefault}{\color[rgb]{1,0,0}cutoff scale}%
}}}}
\put(10576,-2311){\rotatebox{270.0}{\makebox(0,0)[lb]{\smash{{\SetFigFont{17}{20.4}{\familydefault}{\mddefault}{\updefault}{\color[rgb]{0,0,0}adiabatic pressure effects}%
}}}}}
\end{picture}%